\pdfoutput=1
\documentclass[twocolumn,pra,superscriptaddress,longbibliography]{revtex4-2}
\usepackage{graphicx,amsmath,amssymb}
\usepackage[colorlinks=true, citecolor=blue, urlcolor=blue, linkcolor=red]{hyperref}
\begin{document}
\title{Gaussian quantum metrology in a dissipative environment}
\author{Wei Wu}
\affiliation{Lanzhou Center for Theoretical Physics, Key Laboratory of Theoretical Physics of Gansu Province, Lanzhou University, Lanzhou 730000, China}
\author{Jun-Hong An}
\email{anjhong@lzu.edu.cn}
\affiliation{Lanzhou Center for Theoretical Physics, Key Laboratory of Theoretical Physics of Gansu Province, Lanzhou University, Lanzhou 730000, China}

\begin{abstract}
Quantum metrology pursues high-precision measurements of physical quantities by using quantum resources. However, the decoherence generally hinders its performance. Previous work found that the metrological error tends to diverge in the long-encoding-time regime due to the Born-Markovian approximate decoherence, which is called the no-go theorem of noisy quantum metrology. Here we propose a Gaussian quantum metrology scheme using bimodal quantized optical fields as the quantum probe. It achieves the precision of a sub-Heisenberg limit in the ideal case. However, the Markovian decoherence causes the metrological error contributed by the center-of-mass mode of the probe to be divergent. A mechanism to remove this ostensible no-go theorem is found in the non-Markovian dynamics. Our result gives an efficient way to realize high-precision quantum metrology in practical continuous-variable systems.
\end{abstract}
\maketitle

\section{Introduction}

Quantum metrology, which has emerged as a rapidly developing quantum technology, provides a new brand of methodology for realizing high-precision measurements of physical quantities with the help of quantum resources~\cite{PhysRevLett.96.010401,Maccone2011,RevModPhys.90.035006}. It has wide applications from gravitational wave detection~\cite{PhysRevLett.123.231108} and super resolution optical imaging~\cite{Liang:21,doi:10.1063/5.0009538} to quantum thermometries~\cite{PhysRevLett.125.080402,PhysRevResearch.2.033394} and ultra-sensitive magnetometers~\cite{RevModPhys.92.015004,PhysRevApplied.14.044058}. Entanglement is the most common quantum resource to improve the metrology precision. It has been reported that entanglement can be utilized to surpass the so-called shot-noise limit (SNL) \cite{PhysRevLett.79.3865,Nagata726,RevModPhys.90.035005,PhysRevA.94.012339,PhysRevA.102.022602}, which is the fundamental limit set by the laws of classical physics. Such entanglement-enhanced quantum metrology demonstrates its superiority in comparison to its classical counterparts.

Unfortunately, entanglement is very fragile and can be easily destroyed by the decoherence induced by environmental noises. It results in the deterioration of the performance of quantum metrology. It was found that the metrology error generally becomes divergent in the long-encoding-time regime under the influence of decoherence caused by environments ~\cite{Haase_2018,PhysRevA.97.012125,Tamascelli_2020,PhysRevLett.98.160401,PhysRevApplied.5.014007,Albarelli2018restoringheisenberg,PhysRevA.99.033807,PhysRevA.102.012223,PhysRevResearch.2.033389}. Such a phenomenon is called the no-go theorem of noisy quantum metrology \cite{PhysRevLett.116.120801,Albarelli2018restoringheisenberg} and is the main obstacle to achieve a high-precision quantum metrology in practice. However, a clear imperfection leading to this no-go theorem is that it is based on the Born-Markovian approximation to describe the decoherence. Thus, determination of whether this no-go theorem is ostensible or fundamental and can be overcome is highly desirable from both theoretical and experimental perspectives.

To address the above problems, it is necessary to go beyond the Born-Markovian approximation. It has been widely demonstrated that a rigorous non-Markovian treatment gives a qualitatively different dynamical behavior from that of the Born-Markovian approximate case ~\cite{PhysRevLett.109.233601,PhysRevA.88.035806,PhysRevA.102.032607,PhysRevA.81.052330,PhysRevA.93.020105,PhysRevResearch.1.023027,PhysRevA.81.052330,PhysRevA.93.020105,PhysRevResearch.1.023027}. The results inspire us that the non-Markovianity may be exploited to boost the performance of quantum metrology under the influence of practical decoherence.

In this work, we first propose a Gaussian quantum metrology scheme by using a bimodal continuous-variable system as the quantum probe. It achieves a precision as high as the sub-Heisenberg limit (HL) in the ideal case by using Gaussian entanglement as the quantum resource. Then we analyze the effect of decoherence induced by a dissipative environment on the probe. It is found that the decoherence forces the metrology error contributed by the center-of-mass mode to be divergent with increasing encoding time and the superiority to disappear completely in the Born-Markovian approximate dynamics. A mechanism to eliminate this ostensible error-divergence problem as well as enhance the metrology precision within a rigorous non-Markovian dynamical framework is revealed.

\section{Gaussian quantum metrology}

In a typical quantum metrology scheme, one first prepares a quantum probe in a certain state $\rho_{\mathrm{in}}$ and then couples it to the target system such that the measured quantity $\kappa$ is encoded into the probe state $\rho_{\kappa}=\hat{\mathcal{M}}_{\kappa}(\rho_{\mathrm{in}})$. Here, acting on the Liouvillian space of the density matrix, the superoperator $\hat{\mathcal{M}}_{\kappa}$ may be either unitary or nonunitary depending on the measured systems. Finally, one measures a certain observable $\hat{O}$ of the probe in the state $\rho_{\kappa}$ and infers the value $\kappa$ from the results. The inevitable existence of errors results in one being unable to estimate $\kappa$ precisely. According to quantum parameter estimation theory~\cite{Liu_2019}, optimizing all the possible measured observables, the ultimate precision of $\kappa$ is constrained by the quantum Cram\'{e}r-Rao bound $\delta\kappa\geq 1/\sqrt{\upsilon \mathcal{F}_{\kappa}}$, where $\delta\kappa$ is the standard error of $\kappa$, $\upsilon$ is the repeated measurement times, and $\mathcal{F}_{\kappa}\equiv\text{Tr}(\rho_\kappa\hat{L}^2)$, with $\hat{L}$ defined by $\partial_\kappa\rho_{\kappa}=\frac{1}{2}(\hat{L}\rho_\kappa+\rho_\kappa \hat{L})$, is the quantum Fisher information (QFI) characterizing the most information about $\kappa$ extractable from $\rho_{\kappa}$. Due to the independence of $\mathcal{F}_{\kappa}$ on measurement times $\upsilon$, we set $\upsilon=1$ for concreteness in this paper. The performance of quantum metrology is generally characterized by the scaling relation of $\delta\kappa$ or $\mathcal{F}_{\kappa}$ with the number of resource $\bar{n}$ in $\rho_\text{in}$. If $\delta\kappa$ is proportional to $\bar{n}^{-1/2}$ or $\mathcal{F}_{\kappa}\propto\bar{n}$, then the precision is called the SNL. When $\mathcal{F}_{\kappa}\propto\bar{n}^{2}$, it is called the HL. Maximizing the QFI by choosing proper quantum resource in $\rho_{\mathrm{in}}$ and the encoding scheme $\hat{\mathcal{M}}$ is the crucial objective of quantum metrology.

We propose a scheme of quantum metrology by using the entanglement in Gaussian states to enhance the measurement sensitivity. The quantum probe is formed by a bimodal continuous-variable system ~\cite{RevModPhys.77.513,doi:10.1116/5.0007577}. It can be physically realized by two optomechanical cavities~\cite{RevModPhys.86.1391} or nanomechanical resonators~\cite{Forstner:20}. To measure the physical quantity $\kappa$ of a certain classical system, we couple the probe to the system such that the measured quantity $\kappa$ is encoded into the probe state via the dynamics of the probe governed by the Hamiltonian ($\hbar=1$) \cite{PerarnauLlobet2021weaklyinvasive}
\begin{equation}\label{hamilton}
\hat{H}_{\text{p}}=\omega_0\sum_{l=1,2}\hat{a}_{l}^{\dagger}\hat{a}_{l}+\kappa(\hat{a}_1^{\dagger}\hat{a}_2+\text{H.c.}),
\end{equation}
where $\hat{a}_{l}$ is the annihilation operator of the $l$th mode of the probe with the eigen frequency $\omega_0$. The initial state of the probe is chosen to be two-mode squeezed vacuum state $|\Psi(0)\rangle=\mathcal{\hat{S}}_{\hat{a}_{1}\hat{a}_{2}}(r)|{\O}\rangle$ with $\mathcal{\hat{S}}_{\hat{o}}(r)\equiv\exp(r^*\hat{o}-r\hat{o}^{\dagger})$, $r$ being the squeezing parameter, and $|{\O}\rangle$ being the two-mode vacuum state. Such a state contains the mean boson number $\bar{n}=\sum_{l=1,2}\langle\Psi(0)|\hat{a}^\dag_l\hat{a}_l|\Psi(0)\rangle=2\sinh^{2}r$, which can be viewed as the number of quantum resources in our scheme.

It can be found that the initial state $\rho(0)=|\Psi(0)\rangle\langle\Psi(0)|$ is a Gaussian state and such Gaussianity is kept in $\rho(t)=e^{-i\hat{H}_\text{p}t}\rho(0)e^{i\hat{H}_\text{p}t}$ during the time evolution governed by the quadratic Hamiltonian \eqref{hamilton}. The Gaussian state can be described by the characteristic function being of Gaussian form \cite{_afr_nek_2015} $\chi ({\pmb\gamma })=\exp(-{\frac{1}{4}}{\pmb\gamma }^{\dag}{\pmb\sigma}{\pmb\gamma}-i\mathbf{d}^{\dag}\mathbf{K}{\pmb\gamma})$, where ${\pmb \gamma}=(\gamma_{1},\gamma_{2},\gamma_{1}^{*},\gamma_{2}^{*})^{\mathrm{T}}$ and $\mathbf{K}=\mathrm{diag}(1,1,-1,-1)$. The elements of the displacement vector $\mathbf{d}$ and the covariant matrix ${\pmb \sigma}$ are defined as $d_{i} =\text{Tr}[\rho(t) \hat{A}_{i}]$ and $\sigma _{ij} =\text{Tr}[\rho(t) \{\Delta \hat{A}_{i},\Delta \hat{A}_{j}\}]$,
with $\hat{\mathbf A}=(\hat{a}_{1},\hat{a}_{2},\hat{a}_{1}^{\dagger},\hat{a}_{2}^{\dagger})^{\mathrm{T}}$ and $\Delta \hat{A}_{i}=\hat{A}_{i}-d_{i}$. The QFI with respect to the measured quantity $\kappa$ in the bimodal Gaussian state $\rho(t)$ can be calculated via~\cite{_afr_nek_2015,_afr_nek_2018}
\begin{equation}\label{qufsif}
\mathcal{F}_{\kappa}=\frac{1}{2}[\text{vec}(\partial_\kappa {\pmb \sigma})]^\dag{\bf M}^{-1}\text{vec}(\partial_\kappa {\pmb \sigma})+2(\partial_{\kappa}\mathbf{d})^{\dagger}{\pmb\sigma}^{-1}\partial_{\kappa}\bf{d},
\end{equation}
where $\mathbf{M}=\pmb{\sigma}^{*}\otimes{\pmb\sigma}-\mathbf{K}\otimes \mathbf{K}$, with $\pmb{\sigma}^{*}$ being the complex conjugate of $\pmb{\sigma}$. If the state is pure, which causes $\mathbf{M}$ to be noninvertible, then a convenient way to calculate the QFI is
\begin{equation}
\mathcal{F}_{\kappa}=\frac{1}{4}\mathrm{Tr}\big{(}\pmb{\sigma}^{-1}\partial_\kappa {\pmb \sigma}\pmb{\sigma}^{-1}\partial_\kappa {\pmb \sigma}\big{)}+2(\partial_{\kappa}\mathbf{d})^{\dagger}\pmb{\sigma}^{-1}\partial_{\kappa}\mathbf{d}.
\end{equation}

For the sake of convenience, we introduce the center-of-mass and the relative-motion modes $\hat{a}_{\pm}=(\hat{a}_{1}\pm \hat{a}_{2})/\sqrt{2}$. Then we can rewrite Eq. \eqref{hamilton} as $\hat{H}_{\text{p}}=\sum_{\ell=\pm}\omega_{\ell}\hat{a}_{\ell}^{\dagger}\hat{a}_{\ell}$, with $\omega_{\pm}=\omega_{0}\pm\kappa$. One can easily find that, governed by Eq. \eqref{hamilton}, the initial state evolves to $|\Psi(t)\rangle=\otimes_{\ell=\pm}|\psi^{\ell}(t)\rangle$, where $|\psi^{\pm}(t)\rangle=\mathcal{\hat{S}}_{\hat{a}_{\pm}}(\pm re^{-2i\omega_\pm t}/2)|{\O}\rangle$. Due to the independence of the center-of-mass and the relative-motion modes, we have $\mathbf{d}^\text{ideal}=\oplus_{\ell=\pm}\mathbf{d}^\text{ideal}_\pm$ with $\mathbf{d}^\text{ideal}_\pm=(0,0)^\text{T}$ and ${\pmb\sigma}^\text{ideal}=\oplus_{\ell=\pm}{\pmb\sigma}^\text{ideal}_\ell$ with
\begin{equation}
\pmb{\sigma}^\text{ideal}_\pm=\left[
                                \begin{array}{cc}
                                  \cosh(2r) & \mp\sinh(2r)e^{-2i\omega_{\pm}t} \\
                                  \mp\sinh(2r)e^{2i\omega_{\pm}t} & \cosh(2r) \\
                                \end{array}
                              \right].
\end{equation}
According to Eq. \eqref{qufsif}, each pair of $\bf{d}^\text{ideal}_\pm$ and $\pmb{\sigma}^\text{ideal}_\pm$ contributes the QFI as $\mathcal{F}_\kappa^{\pm,\text{ideal}}=2\sinh^{2}(2r)t^{2}$. Then the total QFI can be calculated as
\begin{equation}
\mathcal{F}_{\kappa}^{\mathrm{ideal}}(t)=\sum_{\ell=\pm}\mathcal{F}_\kappa^{\ell,\text{ideal}}=4(\bar{n}^2+2\bar{n})t^2.\label{idqfin}
\end{equation}
It can be seen that the QFI surpasses the HL and thus the achieved metrology precision shows a sub-HL behavior \cite{PhysRevLett.104.103602}. Both the boson number and the encoding time act as the resources to improve the sensitivity.

\section{Effect of dissipative environment}

In practice, the performance of quantum metrology is generally obscured by the presence of ubiquitous decoherence caused by the inevitable interactions of the quantum system with its environment. Depending on whether the system has energy exchange with the environment, the decoherence can be classified into dissipation and dephasing. We consider that the encoding process of the two modes of our Gaussian quantum metrology scheme is disturbed by a common dissipative environment. The Hamiltonian of the total system reads
\begin{equation}\label{haml}
\begin{split}
\hat{H}=&\hat{H}_{\text{p}}+\sum_{k}\Big[\omega_{k}\hat{b}_{k}^{\dagger}\hat{b}_{k}+\sum_{l}\Big{(}g_{lk}\hat{a}_{l}^{\dagger}\hat{b}_{k}+\mathrm{H}.\mathrm{c}.\Big{)}\Big],
\end{split}
\end{equation}
where $\hat{b}_{k}$ denotes the annihilation operator of the $k$th environmental mode with frequency $\omega_{k}$ and $g_{lk}$ is its coupling strength to the $l$th probe mode. We assume that the two modes of the probe homogeneously interact with the environment, which leads to $g_{1k}=g_{2k}=g_{k}$. Then Eq. \eqref{haml} can be rewritten as $\hat{H}=\omega_-\hat{a}_-^\dag\hat{a}_-+\hat{H}_+$ with
\begin{equation}
\hat{H}_+=\omega_{+}\hat{a}_{+}^{\dagger}\hat{a}_{+}+\sum_{k}\Big[\omega_{k}\hat{b}_{k}^{\dagger}\hat{b}_{k}+\sqrt{2}\Big{(}g_{k}\hat{a}_{+}^{\dagger}\hat{b}_{k}+\mathrm{H}.\mathrm{c}.\Big{)}\Big]
\end{equation}
In this situation, only the center-of-mass mode $\hat{a}_+$ feels the presence of the environment, while the relative-motion mode $\hat{a}_-$ is immune to the environment. Commonly, the coupling strength is further characterized by the so-called spectral density $J(\omega)\equiv\sum_{k}|g_{k}|^{2}\delta(\omega-\omega_{k})$. We consider for explicitness that $J(\omega)$ takes the Ohmic-family form \cite{RevModPhys.59.1}
\begin{equation}\label{eq:eq3}
J(\omega)=\eta\omega^{s}\omega_{c}^{1-s}e^{-\omega/\omega_{c}},
\end{equation}
where $\eta$ is a dimensionless coupling constant, $\omega_{c}$ is a cutoff frequency to avoid infrared catastrophe, and $s$ is the so-called Ohmicity parameter. Depending on the value of $s$, the environment can be classified into sub-Ohmic for $0<s<1$, Ohmic for $s=1$, and super-Ohmic for $s>1$.

The environmental presence causes the encoding process to be governed by a nonunitary dynamics of the probe. Considering the environment to be in a vacuum state initially and using the Feynman-Vernon influence-functional method in the coherent-state representation, we can derive an exact non-Markovian master equation for the encoding process as~\cite{PhysRevA.76.042127,qic.9.317}
\begin{eqnarray}\label{eq:eq4}
\dot{\rho}(t)=-i\big{[}\omega_-\hat{a}_-^\dag\hat{a}_-+\Omega(t)\hat{a}_+^{\dagger}\hat{a}_+,\rho(t)\big{]}+\gamma(t)\check{\mathcal{D}}_+\rho(t),~~
\end{eqnarray}
where $\check{\mathcal{D}}_{+}\cdot=2\hat{a}_+\cdot\hat{a}^\dag_{+}-\{\hat{a}^\dag_{+}\hat{a}_+,\cdot\}$, $\Omega(t)=-\text{Im}[\dot{u}(t)/u(t)]$ is the renormalized frequency, and $\gamma(t)=-\text{Re}[\dot{u}(t)/u(t)]$ is the decay rate of the center-of-mass mode of the probe. The time-dependent coefficient $u(t)$ is determined by
\begin{equation}
\dot{u}(t)+i\omega_{+}u(t)+2\int_{0}^{t}d\tau\mu(t-\tau)u(\tau)=0,\label{eomu}
\end{equation}
with $u(0)=1$ and $\mu(x)\equiv\int_{0}^{\infty}d\omega J(\omega)e^{-i\omega x}$.

Solving the master equation \eqref{eq:eq4} under the condition of the initial state being the two-mode squeezed state $|\Psi(0)\rangle$, we can find that the displacement vector and the covariant matrix of $\rho(t)$ are given by $\mathbf{d}(t)=\mathbf{d}_{+}\oplus\mathbf{d}^\text{ideal}_{-}$ and $\pmb{\sigma}(t)=\pmb{\sigma}_{+}\oplus\pmb{\sigma}^\text{ideal}_{-}$, respectively. Here $\mathbf{d}_+=(0,0)^{\mathrm{T}}$ and
\begin{equation}
\pmb{\sigma}_{+}=\left[
                                \begin{array}{cc}
                                  1+2|u(t)|^{2}\sinh^{2}r & -\sinh(2r)u^{2}(t) \\
                                  -\sinh(2r)u^{*2}(t) & 1+2|u(t)|^{2}\sinh^{2}r \\
                                \end{array}
                              \right].\label{covmatx}
\end{equation}The derivation of Eq. \eqref{covmatx} is given in the Appendix.
One can easily check that $\pmb{\sigma}_{+}$ reduces to $\pmb{\sigma}_{+}^{\mathrm{ideal}}$ in the case $\eta=0$. Thus, we have $\mathcal{F}_{\kappa}(t)=\mathcal{F}_{\kappa}^{+}(t)+\mathcal{F}_{\kappa}^{-,\mathrm{ideal}}(t)$, where the QFI with respect to $\{\mathbf{d}_{+},\pmb{\sigma}_{+}\}$ can be calculated via Eq.~(\ref{qufsif}).

In the special case when the system-environment coupling is weak and the characteristic time scale of the environmental correlation function is much smaller than that of the system, one can safely apply the Born-Markovian approximation to Eq.~(\ref{eomu}). The Born-Markovian approximate solution of Eq. \eqref{eomu} reads~\cite{PhysRevA.76.042127,qic.9.317} $u^\text{BMA}(t)\simeq e^{-\{\zeta+i[\omega_{+}+\Delta(\omega_{+})]\}t}$, where $\zeta=2\pi J(\omega_+)$ is the Born-Markovian  approximate decay rate and $\Delta(\omega_{+})=2\mathcal{P}\int _0^\infty {J(\omega)\over \omega_+-\omega}d\omega$, with $\mathcal{P}$ denoting the Cauchy principal value, is the environmentally induced frequency shift. With the approximate expression of $u^\text{BMA}(t)$ at hand, the QFI from $\{\mathbf{d}_{+},\pmb{\sigma}_{+}\}$ in the large-$\bar{n}$ limit reads
\begin{equation}\label{eq:eq13}
\mathcal{F}^{+,\text{BMA}}_{\kappa}(t)\simeq \bar{n}t^{2}[\coth(\zeta t)-1].
\end{equation}
It can be seen from Eq.~(\ref{eq:eq13}) that $\mathcal{F}^{+,\text{BMA}}_{\kappa}(\infty)=0$, which indicates that the metrology error becomes divergent and the corresponding metrology scheme completely breaks down in the long-encoding-time regime. A similar result was also reported in many previous works~\cite{Haase_2018,PhysRevA.97.012125,Tamascelli_2020,PhysRevLett.98.160401,PhysRevApplied.5.014007,Albarelli2018restoringheisenberg,PhysRevA.99.033807,PhysRevA.102.012223,PhysRevResearch.2.033389}. It is understandable based on the fact that the information of $\kappa$ in $\rho(t)$ under the Born-Markovian approximation unidirectionally flows from the probe to the environment such that no message can be extracted in the long-encoding-time regime. On the other hand, via optimizing $t$ in Eq.~(\ref{eq:eq13}), we find that the optimal encoding time is $t=[1+W(-2/e^{2})/2]\zeta^{-1}\simeq0.80\zeta^{-1}$, with $W(x)$ being the Lambert $W$ function. Thus, the corresponding maximal QFI is
\begin{equation}\label{eq:eq14}
\max\mathcal{F}^{+,\text{BMA}}_{\kappa}\simeq 0.32\bar{n}\zeta^{-2},
\end{equation}
which is a SNL-type scaling behavior. In comparison to $\mathcal{F}_{\kappa}^{+,\mathrm{ideal}}$, one can see that the noisy effect of the environment forces the scaling relation from the sub-HL back to the SNL under the Born-Markovian approximation.

In the general non-Markovian case, the expression of $\mathcal{F}_{\kappa}^{+}(t)$ is generally complicated and one must resort to the numerical calculation. However, via analyzing the long-time behavior of $u(t)$, we can obtain an analytical asymptotic form of $\mathcal{F}_{\kappa}^{+}(t)$ in the long-encoding limit. This helps us to create a clear physical picture of the performance of our scheme of Gaussian quantum metrology under the noisy impact of the environment. A Laplace transform to Eq. \eqref{eomu} results in
$\tilde{u}(p)=[p+i\omega_{+}+\int_0^\infty{2 J(\omega)\over p+i\omega}d\omega]^{-1}$. The solution of $u(t)$ is obtained by the inverse Laplace
transform to $\tilde{u}(p)$, which can be exactly done by finding the poles from the transcendental equation
\begin{equation}\label{porot}
Y(E)\equiv\omega_{+}-\int_0^\infty{2 J(\omega)\over\omega-E}d\omega =E,~(E=ip)
\end{equation}
We should point out that the roots of Eq. \eqref{porot} are just the eigenenergies of $\hat{H}_+$ in the single-excitation subspace. To be specific, we expand the eigenstate of $\hat{H}_+$ as $|\Phi\rangle=(x\hat{a}_{+}^{\dagger}+\sum_{k}z_{k}\hat{b}_{k}^{\dagger})|{\O},\{{\O}_k\}\rangle$. Substituting it into $\hat{H}_+|\Phi\rangle=E|\Phi\rangle$, with $E$ being the eigenenergy, we can readily obtain
\begin{eqnarray}
\omega_+x+\sqrt{2}\sum_kg_kz_k&=&Ex,\label{smx}\\
\omega_kz_k+\sqrt{2}g_kx&=&Ez_k.\label{smz}
\end{eqnarray}Substituting the solution of Eq. \eqref{smz} $z_k={\sqrt{2}g_kx\over E-\omega_k}$ into Eq. \eqref{smx}, we obtain
\begin{equation}
\omega_+-2\sum_k{g_k^2\over \omega_k -E}=E. \label{smeig}
\end{equation}Recalling the definition of $J(\omega)$, we readily recover Eq. \eqref{porot} from Eq. \eqref{smeig}. Therefore, on the one hand, Eq. \eqref{porot} completely governs the behavior of $u(t)$ and thus the dynamics of the system and, on the other hand, it gives the eigen-energy of the total system in the single-excitation subspace. This implies that, although the subspaces with higher excitation numbers may be involved in the reduced dynamics, the dynamics of the probe is essentially determined by the single-excitation energy spectrum characteristic of $\hat{H}_+$. Because $Y(E)$ is a monotonically decreasing function in the regime $E<0$, Eq.~(\ref{porot}) has one and only one isolated root $E_b$ in this regime provided $Y(0)<0$. We call the eigenstate corresponding to such isolated eigenenergy $E_b$ a bound state. On the other hand, since $Y(E)$ is not well analytic in the regime $E>0$, Eq.~(\ref{porot}) has infinite roots in this regime, which form a continuous energy band. Then, after applying the inverse Laplace transform and using the residue theorem, we obtain~\cite{PhysRevA.103.L010601}
\begin{equation}\label{eq:eq17}
u(t)=Ze^{-iE_b t}+\int_{0}^{\infty}\frac{2 J(E)e^{-iE t}dE}{[E-\omega_{+}-\Delta(E)]^{2}+[2\pi J(E)]^2},
\end{equation}
where the first term with $Z\equiv[1+\int_0^\infty{2 J(\omega)d\omega\over(E_b-\omega)^2}]^{-1}$ is contributed by the potentially formed bound state and the second term is from the band energies. The second term approaches zero in the long-time regime due to out-of-phase interference of the continuously changing $E$. Thus, if the bound state is absent, we have $u(\infty)= 0$, characterizing a complete decoherence, while if the bound state energy is formed, we have $u(\infty)\simeq Ze^{-iE_b t}$, implying a dissipationless dynamics. The condition, under which the bound state for the Ohmic-family spectral density is formed, can be evaluated as $\omega_{+}-2\eta\omega_c\Gamma(s)\leq 0$, where $\Gamma(s)$ is the Euler Gamma function.

\begin{figure}
\centering
\includegraphics[angle=0,width=8.5cm]{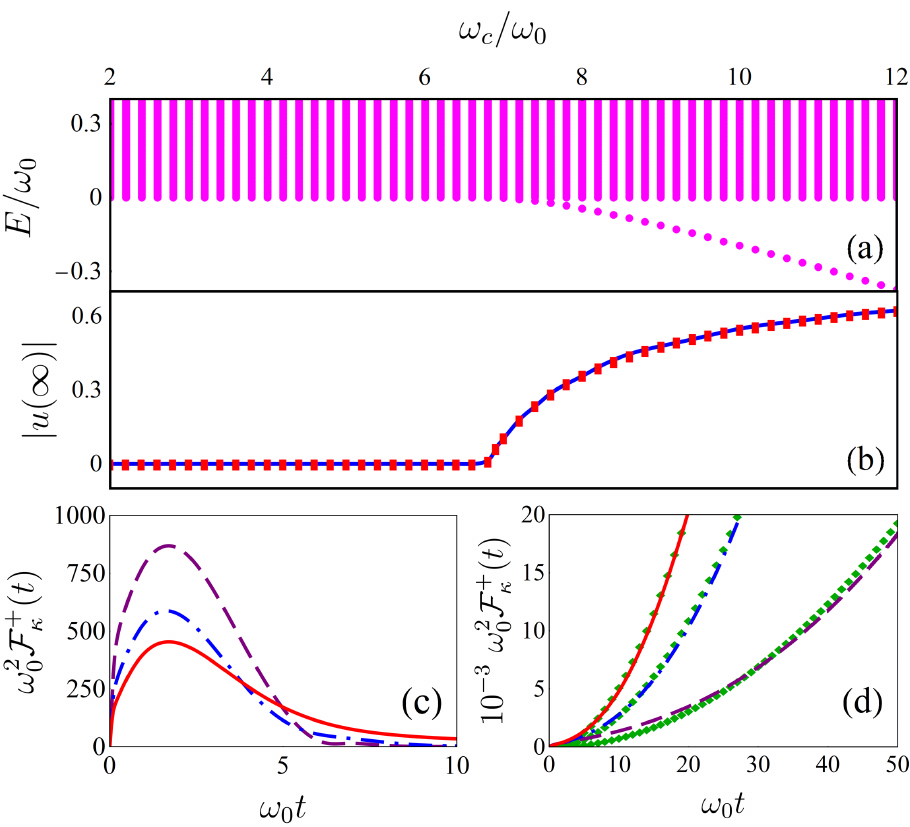}
\caption{(a) Energy spectrum of $\hat{H}$ in the single-excitation subspace. (b) Long-time value of $|u(t)|$ at $t=2\times 10^3\omega_0^{-1}$ (blue solid line) and analytical result of $Z$ (red rectangles). (c) Evolution of $\mathcal{F}_{\kappa}^{+}(t)$ in the absence of the bound state when $\omega_{c}=3\omega_{0}$ (purple dashed line), $4\omega_{0}$ (blue dot-dashed line), and $5\omega_{0}$ (red solid line). (d) Evolution of $\mathcal{F}_{\kappa}^{+}(t)$ in the presence of the bound state when $\omega_{c}=8\omega_{0}$ (purple dashed line), $9\omega_{0}$ (blue dot-dashed line), and $10\omega_{0}$ (red solid line). The green diamonds are analytical results from Eq.~(\ref{eq:eq18}). The other parameters are $\eta=0.05$, $\kappa=0.2\omega_{0}$, $s=0.5$, and $\bar{n}=200$.}\label{fig:fig1}
\end{figure}

\begin{figure}
\centering
\includegraphics[angle=0,width=8.5cm]{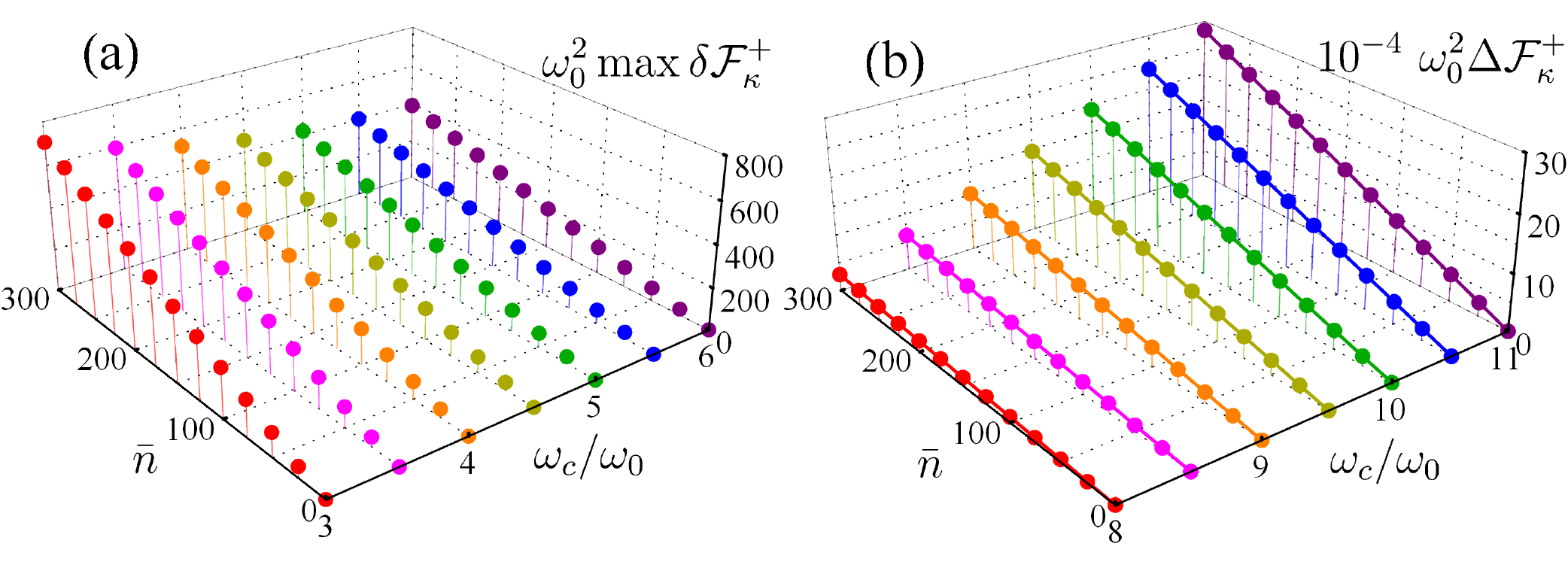}
\caption{(a) Difference between the maximal QFI in the non-Markovian dynamics and that under Born-Markovian approximation, i.e., $\text{max}\delta\mathcal{F}_{\kappa}^{+}=\text{max}\mathcal{F}^+_\kappa(t)-\text{max}\mathcal{F}^{+,\text{BMA}}_\kappa$, as a function of $\bar{n}$ in different $\omega_c$ in the absence of a bound state. (b) Difference between the long-time QFI in the non-Markovian dynamics and the maximal QFI under the Born-Markovian approximation, i.e., $\Delta\mathcal{F}_{\kappa}^{+}=\mathcal{F}^+_\kappa(t=50\omega_0^{-1})-\text{max}\mathcal{F}^{+,\text{BMA}}_\kappa$, as a function of $\bar{n}$ in different $\omega_c$ in the presence of bound state. The dots are numerical results and the lines are analytical results from Eq.~(\ref{eq:eq18}). The other parameters are the same as in Fig.~\ref{fig:fig1}.}\label{fig:fig2}
\end{figure}

It is natural to expect that $\mathcal{F}_{\kappa}^{+}(t)$ asymptotically tends to zero in the absence of the bound state because $u(t)$ approaches zero. This result is qualitatively consistent with the Born-Markovian approximate result, which means our metrology scheme in this case breaks down as well. On the contrary, in the presence of the bound state, substituting the long-time expression of $u(t)$ into Eqs.~(\ref{covmatx}) and \eqref{qufsif}, we have the QFI in the large-$\bar{n}$ limit as
\begin{equation}\label{eq:eq18}
\mathcal{F}^{+}_{\kappa}(t)\simeq \frac{2 Z^{2}}{1-Z^{2}}(\partial_{\kappa}E_{b})^{2}\bar{n}t^{2}.
\end{equation}
It is interesting to see $\mathcal{F}_\kappa^{+}(t)\propto t^{2}$, which means that, quite different from the Born-Markovian approximate and the bound-state-absent cases, the encoding time as a resource to improve the metrology precision is recovered. Such a time scaling relation is the same as the noiseless ideal situation in Eq. \eqref{idqfin}. The result demonstrates that the bound state can be used to retrieve the ideal metrology precision under the noise influence. Such an amazing result is caused by the anomalous equilibrium state induced by the bound state: The message of $\kappa$ is partially preserved in the steady state $\rho(\infty)$ and can be persistently enlarged by prolonging the encoding time. Therefore, we can completely overcome the error-divergence problem appearing in many previous studies~\cite{Haase_2018,PhysRevA.97.012125,Tamascelli_2020,PhysRevLett.98.160401,PhysRevApplied.5.014007,Albarelli2018restoringheisenberg,PhysRevA.99.033807,PhysRevA.102.012223,PhysRevResearch.2.033389} with the help of the bound-state mechanism. Moreover, one can find from Eq.~(\ref{eq:eq18}) that the QFI with respect to $\{\mathbf{d}_{+},\pmb{\sigma}_{+}\}$ behaves as the SNL, i.e., $\mathcal{F}_\kappa^{+}(t)\propto \bar{n}$. Although this result shows the same scaling relation as that of the classical SNL, we still have sufficient room to surpass the SNL and the Born-Markovian approximate one by enhancing the bound-state-favored prefactor in Eq. \eqref{eq:eq18}. It should be noted that the distinguished roles played by the bound state in quantum sensing of a quantum reservoir~\cite{PhysRevApplied.15.054042} and Mach-Zehnder-interferometry-based quantum metrology~\cite{PhysRevLett.123.040402} have been reported.

We now verify our analytical result via numerical calculations. Figure~\ref{fig:fig1}(a) shows the energy spectrum in the single-excitation subspace of the total system. It can be seen that a bound state is formed when $\omega_c>\omega_+/[2\eta\Gamma(s)]$. Accompanying the formation of the bound state, the long-time behavior of $|u(\infty)|$ abruptly increases from zero to a finite value, which exactly matches $Z$ [see Fig.~\ref{fig:fig1}(b)]. Since the dominate role played by $u(t)$ in determining the environmental influence on the encoding dynamics of the probe, this result implies the profound impact of the bound state on our noisy Gaussian quantum metrology. To reveal this clearly, we plot in Figs. \ref{fig:fig1}(c) and \ref{fig:fig1}(d) the evolution of the QFI $\mathcal{F}_{\kappa}^{+}(t)$. It can be seen that $\mathcal{F}_{\kappa}^{+}(t)$, after a transient increase, monotonically decreases to zero in the long-encoding-time limit when the bound state is absent. Thus the encoding time as a resource to enhance the metrology precision in the ideal case \eqref{idqfin} is completely destroyed by the environmental noise effect. This result is qualitatively consistent with the Born-Markovian approximate result in Eq. \eqref{eq:eq13} and many previous studies~\cite{Haase_2018,PhysRevA.97.012125,Tamascelli_2020,PhysRevLett.98.160401,PhysRevApplied.5.014007,Albarelli2018restoringheisenberg,PhysRevA.99.033807,PhysRevA.102.012223,PhysRevResearch.2.033389}. In sharp contrast to this, Fig. \ref{fig:fig1}(d) shows that, as long as the bound state is formed, $\mathcal{F}_{\kappa}^{+}(t)$ exhibits a persistent increase with the encoding time, which matches our analytical result in Eq. \eqref{eq:eq18}. This reveals that the ideal power-law scaling relation \eqref{idqfin} of $\mathcal{F}_{\kappa}^{+}(t)$ with the encoding time is retrieved by the presence of the bound state.

We see from Fig.~\ref{fig:fig1}(c) that maximal QFI exists when the bound state is absent. In order to reveal the non-Markovian effect on the performance of our scheme, we plot in Fig.~\ref{fig:fig2}(a) the difference between this maximal QFI and the Born-Markovian approximate maximal QFI in Eq. \eqref{eq:eq14} as a function of $\bar{n}$ in different $\omega_c$ via optimizing the encoding time. It can be found that the maximal QFI in the non-Markovian dynamics is larger than the Born-Markovian approximate QFI in the full range of parameter. On the other hand, the QFI shows a persistent increasing with the encoding time and no maximal value exists when the bound state is formed. Choosing a large time, we plot the difference between the instantaneous QFI and the maximal QFI in Eq. \eqref{eq:eq14} as a function of $\bar{n}$. Once again, the absolute superiority of the non-Markovian QFI over the Born-Markovian approximate QFI can be observed in the full parameter range.
Therefore, we can conclude that, although both Eqs. \eqref{eq:eq18} and \eqref{eq:eq14} scale with $\bar{n}$ as the SNL, we still have sufficient room to surpass the Born-Markovian approximate QFI not only via increasing the encoding time but also via manipulating the prefactor of Eq. \eqref{eq:eq18} in the non-Markovian dynamics.
This result demonstrates that the non-Markovian effect can boost the metrology precision in the practical noisy case.

\section{Summary}

It should be emphasized that the bound-state-favored superiority in our Gaussian metrology scheme is independent of the explicit form of the spectral density. Although only the Ohmic-family spectral density is displayed in this paper, our result can be generalized to other cases without difficulty. The effects of the non-Markovian effect and bound state have been observed in recent experiments~\cite{PhysRevLett.120.060406,PhysRevLett.124.210502,Liu2016,Krinner2018}, which provides strong support in the experimental realization of our metrology scheme.

In summary, we have proposed a Gaussian quantum metrology scheme by using a two-mode continuous-variable system as the quantum probe. The metrology precision was found to scale with the mean boson number $\bar{n}$ as a sub-Heisenberg limit due to the entanglement in the two-mode squeezed vacuum state of the probe. The decoherence caused by a dissipative environment on the probe was also investigated. It was revealed that the entanglement-favored precision in the relative-motion mode of the probe is immune to the decoherence, while the QFI contributed by the center-of-mass mode tends to vanish with increasing time in the Born-Markovian dynamics. Going beyond this approximation, we further discovered a mechanism to retrieve the encoding time as a metrology resource. It was found that as long as a bound state is formed in the energy spectrum of the total system consisting of the  center-of-mass mode of the probe and the environment, the scaling relation of the QFI with the encoding time in the ideal case is recovered. The mechanism overcomes the error-divergence problem of the center-of-mass mode caused by the decoherence. Sufficiently generalizing the scope of noisy quantum metrology, our result provides an experimentally feasible strategy to realize a high-precision continuous-variable quantum metrology.

\section*{Acknowledgments}

The work was supported by the National Natural Science Foundation (Grants No. 11875150, No. 11834005, and No. 12047501).

\appendix

\section{Derivation of the displacement vector and covariance matrix}\label{appd}

To derive the expressions of $\pmb{\text{d}}_{+}$ and $\pmb{\sigma}_{+}$, we need to calculate $\langle \mathcal{\hat{O}}(t)\rangle\equiv \text{Tr}[\rho(t)\mathcal{\hat{O}}]$, where $\mathcal{\hat{O}}=\hat{a}_{+}$, $\hat{n}_{+}$, and $\hat{a}_{+}^{2}$. From the exact master equation \eqref{eq:eq4}, we have
\begin{eqnarray}
\frac{d}{dt}\langle \mathcal{\hat{O}}(t)\rangle&=&-i\Omega(t)\text{Tr}\Big{[}\hat{a}_+^{\dagger}\hat{a}_+\rho(t)\mathcal{\hat{O}}-\rho(t)\hat{a}_+^{\dagger}\hat{a}_+\mathcal{\hat{O}}\Big{]}\nonumber\\
&&+\gamma(t)\text{Tr}\Big{[}2\hat{a}_{+}\rho(t)\hat{a}_{+}^{\dagger}\mathcal{\hat{O}}-\hat{a}_{+}^{\dagger}\hat{a}_{+}\rho(t)\mathcal{\hat{O}}\nonumber\\
&&-\rho(t)\hat{a}_{+}^{\dagger}\hat{a}_{+}\mathcal{\hat{O}}\Big{]}.
\end{eqnarray}
Then it is straightforward to obtain
\begin{eqnarray}
\frac{d}{dt}\langle \hat{a}_{+}(t)\rangle&=&-[\gamma(t)+i\Omega(t)]\langle \hat{a}_{+}(t)\rangle=\frac{\dot{u}(t)}{u(t)}\langle \hat{a}_{+}(t)\rangle,~~~\\
\frac{d}{dt}\langle \hat{n}_{+}(t)\rangle&=&2\text{Re}\bigg{[}\frac{\dot{u}(t)}{u(t)}\bigg{]}\langle \hat{n}_{+}(t)\rangle,\\
\frac{d}{dt}\langle \hat{a}_{+}^{2}(t)\rangle&=&2\frac{\dot{u}(t)}{u(t)}\langle \hat{a}_{+}^{2}(t)\rangle.
\end{eqnarray}
Solving these differential equations under the initial conditions $\langle \hat{a}_{+}(0)\rangle=0$, $\langle \hat{n}_{+}(0)\rangle=\sinh^2r$, and $\langle \hat{a}_{+}^{2}(0)\rangle=-{\sinh(2r)\over 2}$ for the two-mode squeezed vacuum state, we have
\begin{eqnarray}
\langle \hat{a}_{+}(t)\rangle&=&u(t)\langle \hat{a}_{+}(0)\rangle=0,\\
\langle \hat{n}_{+}(t)\rangle&=&|u(t)|^{2}\langle \hat{n}_{+}(0)\rangle=|u(t)|^{2}\sinh^{2}r,\\
\langle \hat{a}_{+}^{2}(t)\rangle&=&u^{2}(t)\langle \hat{a}_{+}^{2}(0)\rangle=-\frac{1}{2}u^{2}(t)\sinh(2r).
\end{eqnarray}
Then $\pmb{\text{d}}_{+}$ and $\pmb{\sigma}_{+}$ can be obtained.
\bibliography{reference}

\end{document}